------------------------------------------------------------------------------
\documentstyle[12pt]{article}
\newfont{\ssq}{cmss12}
\newfont{\ssmall}{cmss10}
\newfont{\llarge}{cmssbx10 scaled\magstep2}
\newfont{\cbf}{cmssbx10 scaled\magstep1}
\begin{document}
\ssq
\begin{center}
{\llarge SPA+RPA approach to canonical and grandcanonical treatments of
nuclear level densities}\\
\vspace*{0.3 true in}
{\bf B. K. Agrawal$^1$ and P. K. Sahu$^1$}\\
Institute of Physics, Bhubaneswar 751005, India.
\end{center}
\footnotetext[1]{ E-Mail: bijay or pradip@iopb.ernet.in}
\vspace*{1.0 true in}
\begin{abstract}
\ssmall
\noindent Using an exactly solvable pairing model Hamiltonian in the
static path approximation together with small-amplitude quantal
fluctuation corrections in  random phase approximation (SPA+RPA),
we have analyzed the behaviour of canonical (number projected) and
grandcanonical treatments of nuclear level densities as a function of
temperature and  number of particles. For small particle
numbers at a low temperature, we find that though the grandcanonical
partition function in SPA+RPA approach is quite close to its exact
value, the small errors in its estimation causes significant
suppression of level density obtained using number projected
partition function. The results are also compared with the smoothed
out exact values of level density. Within this model study, it appears that
due to saddle point approximation to multiple Laplace-back transform,
the grandcanonical treatment of level density at low temperature may
be reliable only for relatively large number of particles.

\end{abstract}
\newpage

The familiar Bethe  formula\cite{{Bethe pr50},{Bohr 75}} for the
level density,
because of its simplicity, has been widely used to perform the statistical
analysis of nuclear reactions. This level density formula
takes a simple form due to  (i) the connection of grandcanonical
partition function
with microcanonical partition function (level density) by a saddle
point approximation in the evaluation of the traces over the
states\cite{Levit} and (ii) the grandcanonical partition function
itself  is approximated  using independent single particle spectrum which
is further assumed to be  equidistant. Recently, a more
realistic value of the level density\cite{Lauritzen prc39}-\cite{Agrawal
npa576} in saddle point approximation is obtained using a
grandcanonical partition function in
static path approximation\cite{{Alhassid prc30},{Lauritzen prl}}
which accounts for the large-amplitude thermal fluctuations.
The method of  saddle point approximation can be generalized in order to
treat the level densities that are characterized by a set of quantum
numbers, provided, these quantum numbers are composed additively of
contributions from the single particle states (e.g. excitation
energy ($E^*$), number of protons ($N_p$) or neutrons ($N_n$), angular
momentum, parity etc.). For instance, consider the
system consisting of one kind of particles(protons or neutrons). The
exact level density,
\begin{equation}
\label{exact ld}
\rho(E,N)=\sum_{\lambda}\delta(E-E_\lambda)\delta(N-N_\lambda)
\end{equation}
can also be obtained by double  Laplace-back transform of the
grandcanonical partition function ${\cal Z}(\beta,\alpha)$ as
\begin{equation}
\label{lbt}
\rho(E,N)=(\frac{1}{2\pi i})^2 \int \int_{-i\infty}^{+i\infty}
{\cal Z}(\beta,\alpha) e^{\beta E-\alpha N}d\beta d\alpha
\end{equation}
In saddle point approximation, the level density $\hat\rho$ can be obtained
directly in terms of canonical and  grandcanonical partition functions
as
\begin{equation}
\hat\rho^c={Z_N(\beta)e^{\beta E}\over \sqrt{2\pi}
\left[{\partial ^2 ln{\cal Z_{N}}\over \partial\beta^2}\right ]^{1/2} }
\label{saddl}
\end{equation}
and
\begin{equation}
\label{saddle}
\hat\rho^g=
{{\cal Z}(\beta,\alpha)e^{\beta E-\alpha N}\over 2\pi
\left [{\partial^2 ln{\cal Z}\over \partial \beta^2}{\partial^2
ln{\cal Z}\over\partial \alpha^2}-\left ({\partial ^2 ln{\cal Z}\over
\partial\alpha\partial\beta}\right )^2\right ]^{1/2}},
\end{equation}
respectivley.  The canonical partition function $Z_N(\beta)$
is given as
\begin{equation}
Z_N(\beta)={1\over 2\pi i}\int_{-i\infty}^{+i\infty}
Z(\beta,\alpha) e^{-\alpha N} d\alpha
\label{can}
\end{equation}
In above equations, $\beta$ corresponds to inverse temperature ($1/T$) and
$\alpha=\mu\beta$ with $\mu$ being the chemical potential.
Values of $\beta$ and $\alpha$ in eqs. (\ref{saddl}) and (\ref{saddle})
are  chosen such that the saddle point conditions are satisfied,
i.e.,
\begin{equation}
{\partial ln{\cal Z}\over \partial \beta}+E=
{\partial ln{\cal Z}\over\partial \alpha}-N=0.
\end{equation}
{}From eqs. (\ref{saddl}) and (\ref{saddle}), it is clear that in
general the canonical and grandcanonical treatments of level density
would differ by the saddle point approximation to a single and a multiple
Laplace-back transform of corresponding partition functions.
A typical example, the grandcanonical treatment of level density for
a given
nucleus (i.e. fixed  $N_p$ and $N_n$) as a function of
excitation energy and angular momentum would require saddle point
approximation
to four Laplace-back transform\cite{{Agrawal npa576},{Agrawal plb}}.
However, it has been pointed out in ref.\cite{Gross plb318} that the Bethe's
formula
for level density due to the saddle point approximation to several
Laplace-back transform leads to a severe problem $-$  violation of
``microcanonical analyticity constraint" leading to a {\it measurable}
differences between the microcanonical and grandcanonical treatments.
Also, it can be seen in ref.\cite{{Agrawal plb},{Grimes prc42},{Grimes zpa}}
that at very high excitation energy the
assumption  of equidistant
single particle spectrum in Bethe's formula gives
rise to exponential
deviations relative to more realistic values.
\par
In this letter, we  analyze the behaviour of the canonical and
grandcanonical treatments of the level densities. For this
purpose, we have used a recently developed
method of exact functional representation\cite{{Arve anp183},{Lang prc49}}
to construct grandcanonical partition function  for the interacting
systems. It is clear from refs.\cite{{Arve anp183},{Puddu anp}} that
the exact functional
representation can be made easily numerically tractable using
SPA+RPA (static path approximation+random phase approximation)
approach. Moreover, it is shown in refs. \cite{{Puddu anp},{Broglia
prc42},{Quick prc49}} that when SPA+RPA approach applied to  the
grandcanonical partition function for the exactly solvable  pairing
and Lipkin model
Hamiltonians, the results for $E^*$,  $\hat\rho$ and heat capacity
found to  be very close to ones obtained using exact
grandcanonical partition function. This approach has been applied to a
realistic Hamiltonian also\cite{Broglia plb}. In what follows, we
describe briefly, the SPA+RPA approach  to canonical (number
projected) and
grandcanonical partition functions for exactly solvable pairing model.

\par We consider here the pairing Hamiltonian defined as
\begin{equation}
\hat H=-G\hat P^\dagger \hat P
\label{ph}
\end{equation}
where $G$ is the interaction strength, $\hat
P=\sum_{k>0}\hat a_{\scriptscriptstyle \bar k}\hat
a_{\scriptscriptstyle { k}}$.
The index $k=1,2,...\Omega$
with $\bar k=-k$ labels $2\Omega$ degenerate single particle states.
Equation (\ref{ph}) can be rewritten in terms of quasi-spin operators
as
\begin{equation}
\hat H=-G(\hat J^2-\hat J_z^2+\hat J_z)
\label{ph qs}
\end{equation}
where, $\hat J_x={1\over 2}(\hat P+\hat P^\dagger)$, $J_y={i\over 2}(\hat
P-\hat P^\dagger)$ and $\hat J_z={1\over 2}(\hat N-\Omega)$. Since,
$\left [\hat H,\hat J^2\right ]=\left [\hat H,\hat J_z\right ]=0$,
the eigenvalues can be given as
\begin{equation}
E(J,M)=-G(J(J+1)-M^2+M)
\label{eigen}
\end{equation}
where,
\begin{equation}
\label{m}
M={1\over 2}(N-\Omega)
\end{equation}
 with $N$ being the
number of particles. The degeneracy factor for a given $J$ state  is
\begin{eqnarray}
\label{gj}
g_{\scriptscriptstyle J}=
\left (
\begin{array}{c}
2\Omega\\
\Omega-2J
\end{array}
\right )
-
\left (
\begin{array}{c}
2\Omega\\
\Omega-2J-2
\end{array}
\right ).
\end{eqnarray}
With these ingredients (eqs. \ref{eigen}-\ref{gj}), the exact level
density can be obtained using eq. (1) as
\begin{equation}
\rho(E,N)=\sum_{\scriptscriptstyle J}g_{\scriptscriptstyle
J}\delta(E-E(J)).
\end{equation}
Here we have suppressed the $M$ dependence of eigenvalues (see
eq.(\ref{eigen})) since it is fixed for a given number of particles.
We define now a smooth version as
\begin{equation}
{\bar\rho(E,N)}=\sum_{\scriptscriptstyle J}g_{\scriptscriptstyle
J}\Delta(E-E(J))
\label{smooth}
\end{equation}
which describes the exact average level density. We adopt a simple
form for  the
spreading function $\Delta(x)$ as given below,
\begin{equation}
\Delta^n(x)={n\over \sqrt{\pi}}e^{-n^2x^2}
\end{equation}
where the parameter $n$ controls the width of spreading function.
The above choice of $\Delta$-function essentially represents the
dominant part of $\Delta=L^{1/2}_3 e^{-x^2}$ as used in ref.
\cite{Alhassid npa549}.
\par It follows immediately from   eqs.
(\ref{saddl})-(\ref{can}) that the level density in saddle point
approximation can easily be obtained once  the partition function is
known. The exact canonical or
grandcanonical partition function can be constructed as
\begin{equation}
\label{cexact}
{\cal Z}_N(\beta)=\sum_{\scriptstyle J}g_{\scriptscriptstyle J} e^{-\beta
E_J}
\end{equation}
and
\begin{equation}
{\cal Z}(\beta,\alpha)=\sum_Ne^{\alpha N}\sum_{\scriptstyle J}
g_{\scriptscriptstyle J}e^{-\beta E_J}
\label{gexact}
\end{equation}
where $J=\mid M\mid, \mid M\mid +1,...,\Omega/2$ or $(\Omega-1)/2$
depending on whether the $N$ is even or odd. It is worth mentioning
here that eq. (\ref{cexact}) can precisely be obtained by substituting eq.
(\ref{gexact}) in eq. (\ref{can}) and using ``Wick
rotation"(i.e., $\alpha\rightarrow i\alpha$).
\par
In SPA+RPA approach, ${\cal Z}(\beta,\alpha)$ takes the following form,
\begin{equation}
{\tilde{\cal Z}} (\beta,\alpha)={2\beta\over G}\int_0^\infty  \Delta
d\Delta e^{-{\beta\over G}
\Delta^2+\alpha\Omega+ \Omega S+2\Omega ln[1+e^{-S}]}{\cal C}(\Delta)
\label{Z}
\end{equation}
where, the factor ${\cal C}(\Delta)$ as given below
\begin{eqnarray}
{\cal C}(\Delta)=\prod_{m>0}\left[\left [1-{\beta\over 2}G\Omega {Q^2\over S}
{tan{S\over 2}\over S^2+(\pi m)^2}  \right ]\left [1-{\beta\over
2}G\Omega S {tan{S\over 2}\over S^2+(\pi m)^2}  \right ] \right  .
\nonumber \\
\left .  + \left [{\beta\over 2}G\Omega {Q\over S}
{(\pi m) tan{S\over 2}\over S^2+(\pi m)^2}  \right ]^2 \right ]^{-1}
\end{eqnarray}
is due to RPA, which accounts for small-amplitude quantal fluctuation
corrections and
$Q=\alpha+{\beta\over 2}G,~ S=\sqrt{Q^2+\beta^2\Delta^2}$.
With ${\cal C}=1$, eq. (\ref{Z}) reduces to the partition
function in SPA approach. The superscript tilde on ${\cal Z}$ in
above equation is used to distinguish it from exact one.

\par The canonical (number projected) partition function can be obtained as
follows
\begin{equation}
{\tilde{\cal Z}}_N(\beta)={1\over
2\pi}\int_{-\infty}^{+\infty}d\alpha e^{-i\alpha
N}{\tilde{\cal Z}}(\beta,i\alpha)
\label{crpa}
\end{equation}
where ${\tilde{\cal Z}}(\beta,i\alpha)$ is obtained using eq. (\ref{Z}) with
$\alpha \rightarrow i\alpha$. The symmetric interval in $\alpha$
ensures that the partition function ${\cal Z}_N(\beta)$
will be real even if the
approximate form for grand partition function is used.
\par
We analyze now the saddle point approximation to the level density
$\hat\rho$, evaluated using (a) SPA+RPA to canonical (number
projected), (b) SPA+RPA to grandcanonical, (c)
exact canonical and (d) exact grandcanonical partition functions. For
the sake of comparison of these results with the smoothed out exact
value of level density (eq. \ref{smooth}), we have plotted in figures
\ref{fig1} and \ref{fig2} the ratio $\hat\rho/\bar\rho$ as a function of
temperature for N = 5 and 10, respectively. To facilitate the further
discussions, we denote $\hat \rho$ for the cases (a), (b), (c) and
(d) by $\hat\rho^c_{\scriptstyle {rpa}}$, $\hat\rho^g_{\scriptstyle {rpa}}$,
$\hat\rho^c_{\scriptstyle {ex}}$ and $\hat\rho^g_{\scriptstyle
{ex}}$, respectively.

\par
It can be  seen from figures \ref{fig1} and \ref{fig2} that for all
the cases,
$\hat\rho/\bar\rho$ is quite different from unity at low temperature
($T\approx 0.25$ MeV) but with increase in temperature it approaches unity.
The negative slope in $\hat\rho/\bar\rho$ vs $T$ curves
indicates that the saddle point approximation is not applicable at
very low temperature ($T<$0.25 MeV). So, we restrict ourselves to
$T\ge 0.25$ MeV. Practically,  for $N=10$, the difference between
$\hat\rho^g_{\scriptstyle rpa}/\bar\rho$ and
$\hat\rho^g_{\scriptstyle ex}/\bar\rho$
is insignificant.
On the other hand, for $N$=5 the difference between
$\hat\rho^c_{\scriptstyle rpa}/\bar\rho$ and
$\hat\rho^c_{\scriptstyle ex}/\bar\rho$ is noticeable at low $T$,
whereas, this difference is small for $N$=10.
It implies that, though the grandcanonical
partition
function in SPA+RPA approach becomes almost exact, the small errors in
its estimation causes a significant suppression of the level density
obtained using number projected partition function (eq. \ref{crpa})
 at low $T$. Similar behaviour
has been observed by Walt and Quick \cite{Quick prc49} for the heat
capacity. In other words, at low temperature, the small errors in
grandcanonical partition
function enhances the difference between ${\tilde{\cal Z}}_N(\beta)$ and
${\cal Z}_N(\beta)$.

\par
Let us now analyze the difference between canonical and grandcanonical
treatments of level density and their deviations from the exact values.
As discussed above (eqs. \ref{saddl} and \ref{saddle}), in the present
considerations, these
treatments would differ only by the saddle point approximation to a
single and a double Laplace-back transform. It is obvious from
figures \ref{fig1} and \ref{fig2}, the relative difference between the curves
(a) and (b) or (c) and (d) at a fixed $T$ and $N$ is proportional to
$\hat\rho^c- \hat\rho^g$. The difference between the level densities using
canonical and grandcanonical treatments decreases with increase in
temperature or number of particles. For instance at T = 0.35 MeV, the
ratio $\hat\rho^c_{\scriptstyle {ex}}/\hat\rho^g_{\scriptstyle {ex}}$ =
20.83 and 0.94 for N = 5 and 10, respectively, whereas, for T $>$ 0.45
MeV this ratio becomes nearly equal to one.
Finally, coming to the comparison between $\hat\rho$ and $\bar\rho$, we
find that $\hat\rho/\bar\rho$ tends to unity with increase in number
of particle and temperature.
Thus, we can say that the saddle point approximation becomes almost
exact with increase in temperature and particle numbers.
\par
In conclusion, we have analyzed the canonical and grandcanonical
treatments of the nuclear level densities in SPA+RPA approach which
accounts for large-amplitude thermal fluctuations and small-amplitude
quantal fluctuations. The results are compared with the
corresponding exact ones. It is found that at low temperature
the small errors in estimating the grandcanonical partition function
cause significant suppression of level density obtained using number
projected (canonical) partition function for small particle numbers.
Also, we found that due to
saddle point approximation to multiple Laplace-back transform, the
grandcanonical treatment of the level density would be reliable at
low temperature provided the particle number is relatively large.
Furthermore, we must say that except at low temperatures where
saddle point approximation itself is not reliable, a realistic value
of level density can be obtained using SPA+RPA representation for the
grandcanonical partition function.

\vspace*{0.5 true in}
\noindent {\llarge Acknowledgement}
\par
We would like to thank A. Ansari for critical reading of the
manuscript and instructive suggestions.
\newpage
\begin{figure}
\noindent {\llarge Figure Captions}
\caption{\label{fig1} Plot for $\hat\rho/\bar\rho$ vs
temperature for $N=5$,
where $\bar\rho$ represents the smoothed out exact (microcanonical)
value of the
level density and $\hat\rho$ correspond to the level density in a
saddle point approximation which  is obtained using (a) SPA+RPA
approach to the canonical, (b) SPA+RPA approach to the grandcanonical,
(c) exact canonical and (d) exact grandcanonical partition functions.}

\caption{\label{fig2}  Same as figure 1 but for N = 10. Note that the
curves (a) and (b) or (c) and (d) are very close to each other at all
temperatures. }
\end{figure}
\vspace*{4 true in}

\end{document}